\documentclass[10pt,aps,prd,twocolumn,floats,floatfix,showpacs,superscriptaddress,nofootinbib]{revtex4-2}
\usepackage[utf8]{inputenc}               		

\usepackage{graphicx,mathtools,amssymb,amsmath,amsthm,amsfonts,epsfig,epsf}
\usepackage[outdir=./]{epstopdf}
\usepackage[dvipsnames]{xcolor}
\usepackage[linktocpage]{hyperref}
\hypersetup{
    colorlinks=true,
    linkcolor=blue,
    filecolor=magenta,      
    urlcolor=BrickRed,
    citecolor=BrickRed,
}
\hypersetup{pdfstartview=}	
\usepackage{tensor}	
\usepackage{csquotes}
\usepackage{mathrsfs}

\usepackage{graphicx}
\usepackage{amsmath}
\usepackage{autobreak}

\newcommand{\dd}{\mathrm{d}}

\newcommand{\gb}{\mathcal{G}}

\newcommand{\meff}{m_{\text{eff}}}
\newcommand{\beq}{\begin{equation}}
\newcommand{\eeq}{\end{equation}}

\begin{document}

\allowdisplaybreaks

\title{Stable spontaneously-scalarized black holes in generalized scalar-tensor theories}

\author{Georgios~Antoniou}
\affiliation{Nottingham Centre of Gravity,
Nottingham NG7 2RD, United Kingdom}
\affiliation{School of Mathematical Sciences, University of Nottingham,
University Park, Nottingham NG7 2RD, United Kingdom}

\author{Caio F. B. Macedo}
\affiliation{Faculdade de Física, Universidade Federal do Pará, Salinópolis, Pará, 68721-000 Brazil} 

\author{Ryan~McManus}
\affiliation{Nottingham Centre of Gravity,
Nottingham NG7 2RD, United Kingdom}
\affiliation{School of Mathematical Sciences, University of Nottingham,
University Park, Nottingham NG7 2RD, United Kingdom}
\affiliation{School of Physics and Astronomy, University of Nottingham,
University Park, Nottingham NG7 2RD, United Kingdom}

\author{Thomas~P.~Sotiriou}
\affiliation{Nottingham Centre of Gravity, Nottingham NG7 2RD, United Kingdom}
\affiliation{School of Mathematical Sciences, University of Nottingham,
University Park, Nottingham NG7 2RD, United Kingdom}
\affiliation{School of Physics and Astronomy, University of Nottingham,
University Park, Nottingham NG7 2RD, United Kingdom}

\begin{abstract}
It has been shown that the synergy of a scalar field coupling with both the Ricci scalar and the Gauss-Bonnet invariant significantly affects the properties of scalarized black holes and neutron stars, including their domain of existence and the amount of scalar hair they carry. 
Here we study the radial stability of scalarized black-hole solutions.
We demonstrate that they are stable against radial perturbations for Ricci couplings consistent with both a late-time cosmological attractor and the evasion of binary pulsar constraints. 
In addition, we investigate the effect of the Ricci coupling on the hyperbolicity of the equation governing linear, radial perturbations and show that it significantly reduces the region over which hyperbolicity is lost.
\end{abstract}

\maketitle

\section{Introduction}
The first direct detection of gravitational waves in 2015 \cite{LIGOScientific:2016aoc} by the LIGO-Virgo collaboration signaled the start of a new era in astrophysics. 
Observations of gravitational waves generated by
merging compact objects, namely black holes (BH) and neutron stars (NS), provide a powerful tool to probe the strong field, dynamical regime of gravity for the first time in history. 
Despite the success of General Relativity (GR) in the weak field limit, deviations from GR could become relevant when gravitational effects are extreme. 
Indeed, gravitational-wave observations have already been used to place constraints on theories that seek to modify GR and look beyond the Standard-Model framework \cite{Yunes:2016jcc,Dima:2017pwp,LIGOScientific:2018dkp,Ma:2019rei,LIGOScientific:2020tif}.

Considering the overwhelming success of GR in the weak-field regime, noteworthy theories are those which exhibit novel gravitational effects that only manifest in the vicinity of black holes and compact neutron stars but lie dormant elsewhere. To that end, \textit{spontaneous scalarization} is a strong gravity effect in which a phase transition endows black holes and neutron stars with an extra scalar configuration. 
The phase transition is induced by a coupling of the scalar field with curvature invariants, and the details of this coupling control the transition threshold. At a linear level, such couplings trigger a tachyonic instability, and consequently the development of scalar hair. 
The phenomenon was originally identified in Ref.~\cite{Damour:1993hw} for a class of theories where it only takes place for compact stars. 
However, these theories are unable to induce scalarization on black hole spacetimes as they are subject to the no-hair theorems, \cite{PhysRevLett.28.452,Hawking:1972qk,Sotiriou:2011dz,Sotiriou:2015pka}, and so cannot support non-trivial scalar profiles.
Recently, it been shown that another class of theories can exhibit black hole scalarization, \cite{Doneva:2017bvd,Silva:2017uqg}, by including a coupling with the Gauss-Bonnet (GB) invariant, ${\cal G}= R^{\mu\nu\lambda\sigma}R_{\mu\nu\lambda\sigma}-4R^{\mu\nu}R_{\mu\nu} +R^2$.
Here $R^{\mu}_{\nu\lambda\sigma}$ denotes the Riemann tensor, $R_{\mu\nu}$ the Ricci tensor, and $R$ the Ricci scalar. 

In particular, Ref.~\cite{Silva:2017uqg} considered the action
\begin{equation}
\label{eq:ActionsGB}
\begin{split}
    S=\frac{1}{2\kappa}\int \dd^4x\sqrt{-g}\bigg[& R + X +f(\phi) \gb  \bigg],
\end{split}
\end{equation}
where $\kappa = 8\pi G/c^4$, $X=-(\partial\phi)^2/2$, and showed that stationary and asymptotically flat black holes are described by the Kerr metric provided $f'(\phi_0)=0$ for some constant $\phi_0$ and $f''(\phi_0) \gb<0$. 
Further, it was found that $-f''(\phi_0) \gb$ acts as an effective mass squared for the scalar perturbation; hence, when $f''(\phi_0) \gb$ becomes positive and sufficiently large it triggers a tachyonic instability. 
A $f(\phi)\propto \phi^2$ coupling was considered, being the simplest model that satisfied the conditions for developing the instability.
Indeed, scalarization was found to occur and the scalarized black hole solutions exist in the same region of parameter space as the tachyonic instability. 
Ref.~\cite{Doneva:2017bvd} focused instead on the case where $f(\phi)\propto e^{\phi^2}$, also showing that tachyonic instabilities occur and that scalarized black holes again exist in the same region of the parameter space. 
 
Extensions of these models have examined the effects of additional terms in the action, including a bare mass and higher order corrections, different fields, or different types of instabilities as triggers of scalarization, {\em e.g.}~\cite{Ramazanoglu:2016kul,Blazquez-Salcedo:2018jnn,Silva:2018qhn,Macedo:2019sem,Herdeiro:2018wub,Ramazanoglu:2017xbl,Ramazanoglu:2018hwk}. 
In more recent work, it has been shown that a tachyonic instability leading to scalarization can be triggered by spin \cite{Dima:2020yac} and spacetimes describing such scalarized black holes have been generated \cite{Herdeiro:2020wei,Berti:2020kgk}. 
The onset of scalarization is controlled by terms that contribute to linear perturbations around a GR background in all of these scenarios. 
The minimal action that contains all such terms for scalar-tensor theories, up to field redefinitions, and leads to second order equations upon variation was identified in Ref.~\cite{Andreou:2019ikc}:
\begin{equation}
\label{eq:ActionGeneric}
\begin{split}
    S=\frac{1}{2\kappa}\int \dd^4x\sqrt{-g}\bigg[& R + X +\gamma\, G^{\mu\nu}\nabla_\mu \phi \nabla_\nu \phi
    \\
    &-\left(m_\phi^2+\frac{\beta}{2}R-\alpha \gb \right)\frac{\phi^2}{2} \bigg],
\end{split}
\end{equation}
where $\kappa = 8\pi G/c^4$, $X=-(\partial\phi)^2/2$, $m_\phi$ is the bare mass of the scalar field, $\beta$ is a dimensionless parameter, and $\alpha$ has dimensions of length squared. 
Since vacuum spacetimes in GR satisfy $G_{\mu\nu}=0$, the derivative coupling to $G_{\mu\nu}$ and the coupling between the scalar and $R$ will not contribute to the onset of the tachyonic instability for black holes. 
These terms will however affect scalarization thresholds for neutron stars \cite{Ventagli:2020rnx} and also affect the end state of scalarization for both neutron stars \cite{Ventagli:2021ubn} and black holes \cite{Antoniou:2021zoy}. 
Indeed, the tachyonic instability is quenched by non-linearities.
Therefore, these terms affect the final properties of the scalarized configurations, as do any other terms that have been neglected in action~\eqref{eq:ActionGeneric} because they do not contribute to the linear perturbation equation. 

Understanding how the various (self)interactions beyond the scalar-Gauss-Bonnet coupling affect the scalar profile of a scalarized compact object is essential from an observational perspective.
It has been shown that scalarized black holes are unstable under radial perturbations in the simplest, quadratic coupling scenario \cite{Blazquez-Salcedo:2018jnn}.
This issue can be overcome if one considers an additional quartic interaction in the Gauss-Bonnet coupling function, provided that the sign of the quartic coupling coefficient is opposite to the quadratic one \cite{Silva:2018qhn}. 
However, addressing the instability with quartic, or exponential couplings is not entirely appealing from an effective field theory (EFT) perspective. 
This is because these terms have a higher mass dimension than other terms that could contribute non-linearly, e.g. a simple $\phi^4$ self interaction.
It was, indeed, shown in Ref.~\cite{Macedo:2019sem} that including self interactions for the scalar can lead to radially stable scalarized solutions.

More recently, Ref.~\cite{Antoniou:2021zoy} provided strong indications that a coupling of the scalar field with the Ricci scalar, already present in action \eqref{eq:ActionGeneric}, can also lead to radially stable solutions. 
This would be particularly interesting if proven to be true as this same coupling has already been shown to address observational viability issues for black hole scalarization models when $\beta>0$: it dominates the scalar dynamics in the late-time cosmology and turns GR into a cosmological attractor \cite{Antoniou:2020nax}; it can also suppress scalarization of neutron stars \cite{Ventagli:2021ubn} and hence, remove binary pulsar constraints.  
In this paper, we perform a radial perturbation analysis for scalarized solutions and fully explore the role of the coupling with the Ricci scalar in the stability of the solutions.

The structure of the paper is as follows: in Sec.~\ref{sec:setup} we, first, introduce our model and present the field equations. We reproduce the results already derived in previous work, concerning the background BH solutions, that describe static and spherically symmetric configurations. Then, we consider radial perturbations to the static background and discuss the properties of our model as a Schr\"odinger-like problem. In Sec.~\ref{sec:numerical_results} we discuss our numerical results regarding the stability of the solutions and the hyperbolicity of the scalar perturbation equation. Finally, in Sec.~\ref{sec:conclusions} we present our conclusions.

\section{Setup}
\label{sec:setup}

\subsection{Action and field equations}

We will consider the following action 
\begin{equation}
\label{eq:Action}
\begin{split}
    S=\frac{1}{2\kappa}\int \dd^4x\sqrt{-g}\bigg[& R + X 
    -\left(\frac{\beta}{2}R-\alpha \gb \right)\frac{\phi^2}{2} \bigg],
\end{split}
\end{equation}
motivated in part by simplicity and in part by the fact that we seek to understand the influence of the coupling between the scalar and $R$ in stability considerations. One can think of this action as part of an EFT in which the scalar enjoys $\mathbb{Z}_2$ symmetry and shift symmetry is broken by the couplings to curvature.
Note that the model considered in  Refs.~\cite{Damour:1992kf,Damour:1993hw}  is equivalent to action \eqref{eq:Action} with $\alpha=0$ in linear perturbation theory \cite{Andreou:2019ikc} and our definition of $\beta$ matches that of Refs.~\cite{Damour:1992kf,Damour:1993hw}.

The field equations for the metric that one derives from action \eqref{eq:Action} are
\begin{equation}\label{eq:grav_eq}
    G_{\mu\nu}=T^\phi_{\mu\nu},
\end{equation}
where 
\begin{equation}
\label{eq:EMtensor}
\begin{split}
    T^\phi_{\mu\nu}=&-\frac{1}{4}g_{\mu\nu}(\nabla\phi)^2+\frac{1}{2}\nabla_\mu\phi\nabla_\nu\phi\\
    &+\frac{\beta\phi^2}{4}G_{\mu\nu}+\frac{\beta}{4}\left(g_{\mu\nu} \nabla^2 -\nabla_\mu\nabla_\nu \right)\phi^2\\
    &-\frac{\alpha}{2 g}g_{\mu(\rho}g_{\sigma)\nu}\epsilon^{\kappa\rho\alpha\beta}\epsilon^{\sigma\gamma\lambda\tau}R_{\lambda\tau\alpha\beta}\nabla_{\gamma}\nabla_{\kappa}\phi^2
\end{split}
\end{equation}
The scalar field equation reads
\begin{equation}\label{eq:scal_eq}
\Box \phi =m_\text{eff}^2\,\phi,
\end{equation}
where the effective scalar mass is given by
\begin{equation}
    \meff^2=\frac{\beta}{2}R-\alpha \gb.
\end{equation}

\subsection{Spherical Black Hole Solutions}
The background considered is static and spherically symmetric, and described by the metric
\begin{equation}
    ds^2=-A(r)dt^2+\frac{1}{B(r)}dr^2+r^2 d\Omega^2,
\end{equation}
with a scalar field that depends only on the radial coordinate, $\phi=\phi(r)$. 
The differential equations describing the unknown functions $(A(r),B(r),\phi(r))$ can be obtained from \eqref{eq:grav_eq} and \eqref{eq:scal_eq} which are explicitly shown in Appendix~\ref{app:equations}.

In \cite{Antoniou:2021zoy} BH solutions that are asymptotically flat, and continuously connected to GR were found and their characteristics were studied.
From these solutions one can infer the ADM mass $M$ and scalar charge $Q$ from the large $r$ behaviour of $B(r)$ and $\phi$
\begin{align}
    g_{rr}\xrightarrow{r\to \infty} &\; 1-2M/r\,,\\
    \phi\xrightarrow{r\to \infty} &\; Q/r.
\end{align}
For the expansions including terms up to $\mathcal{O}(r^{-6})$ refer to \cite{Antoniou:2021zoy}.
The solutions depend on the scaling of the BH mass and scalar charge with respect to the parameter $\alpha$, controlling the GB coupling. 
When analyzing the stability of the solutions, a dimensionless re-parametrization proves particularly useful:
\begin{equation}
    \hat{M}\equiv {M}/{\alpha^{1/2}} ~{\rm and} ~\hat{Q}\equiv  {Q}/{\alpha^{1/2}}.
\end{equation}

The domain of existence for black holes with non-zero scalar charge is non-trivial. 
To analyze the parameter space in which these solutions exist, we scan $(\alpha/r_h^2,\beta)$ for BHs with a non-trivial scalar field configuration that vanishes asymptotically, in accordance with \cite{Antoniou:2021zoy}. 
For $\alpha=0$ the only solution that is regular at the horizon and asymptotically flat is the Schwarzschild BH~\cite{Sotiriou:2011dz}, while for $\beta=0$, the allowed values for $\alpha$ are in agreement with \cite{Doneva:2017bvd,Silva:2017uqg}

\begin{figure*}[!tbp]
  \centering
  \includegraphics[width=0.32\textwidth]{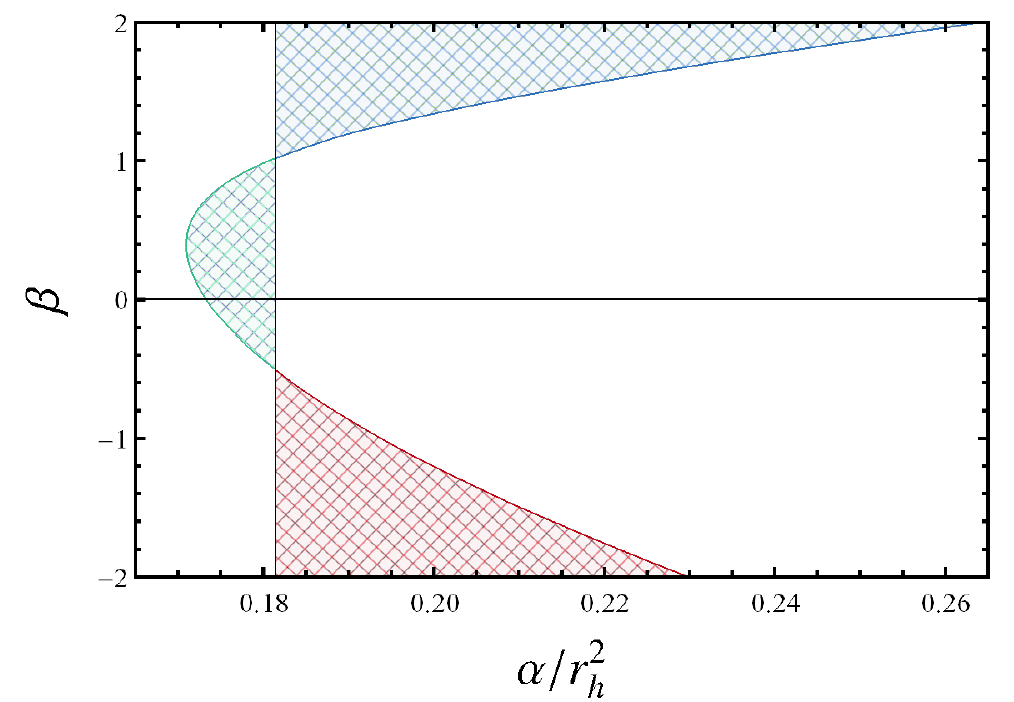}
\hspace{1mm}
  \includegraphics[width=0.32\textwidth]{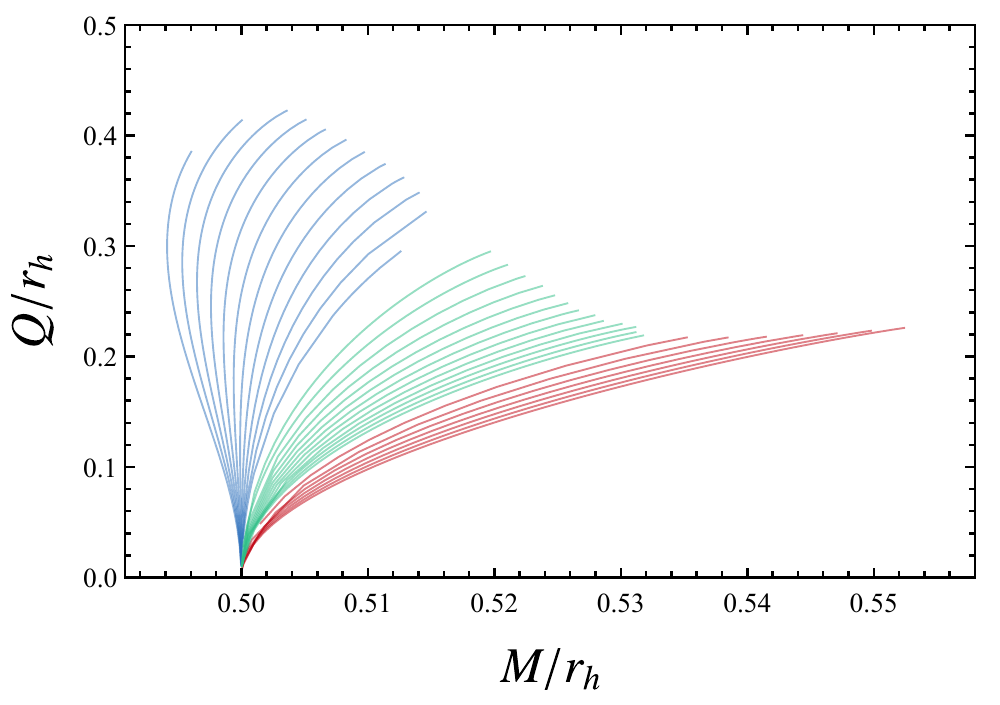}
  \hspace{1mm}
  \includegraphics[width=0.32\textwidth]{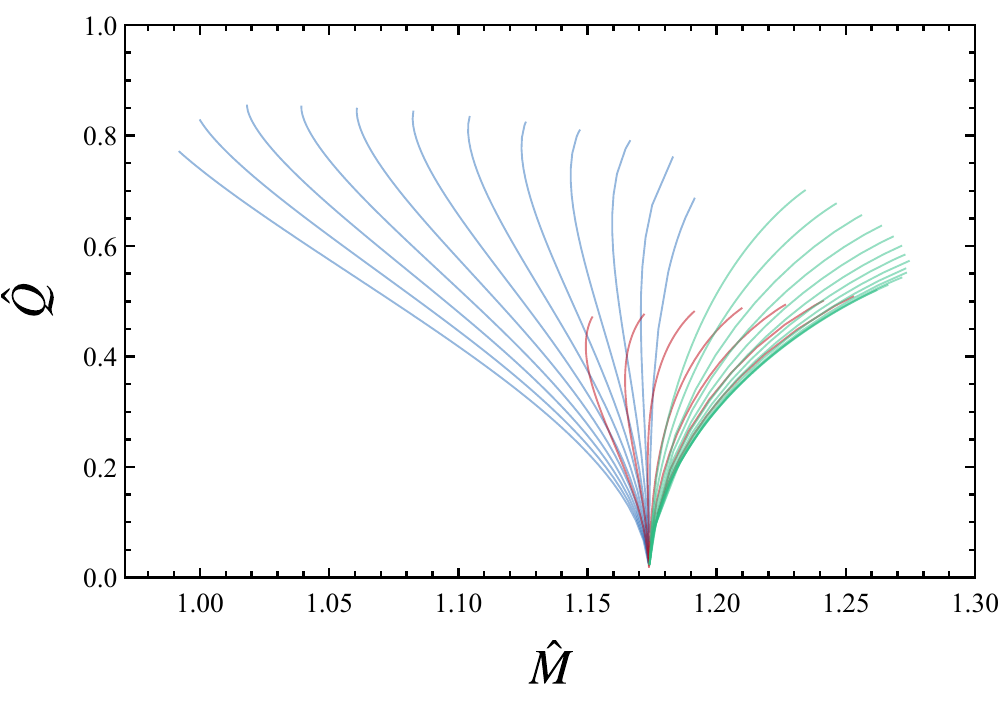}
  \caption{
  \textit{Left:}
  The shaded part corresponds to the region in the parameter space $(\beta,\alpha/r_h^2)$ where scalarized, and asymptotically flat black hole solutions exist. \textit{Centre:} The mass and charge for scalarized black holes.
  The blue, green and red regions correspond to the upper, middle and lower regions of the left-panel plot respectively. The contours shown are of constant $\beta$ with separation of $\delta\beta=0.04,~0.01,~0.24$ in the blue, green and red regions respectively. \textit{Right:} Same as the centre plot but for mass and charge normalised with respect to $\alpha$.
  }
  \label{fig:domains}
\end{figure*}

The left panel of Fig.~\ref{fig:domains} shows the existence domain in the $(\alpha/r_h^2,\beta)$ plane. 
Scalarized solutions with the desired properties, {\em i.e.}~regular everywhere and asymptotically vanishing, exist in the shaded regions.
We can classify the space of solutions existing in three domains bounded by a seemingly vertical line given by the critical value for scalarization  $\alpha/r_h^2\approx0.18$ and a parabola-like curve defined by the existence condition presented in Ref.~\cite{Antoniou:2021zoy}.

To further analyze the solutions, in the center panel of Fig.~\ref{fig:domains}, we show the domain of existence in the $(M/r_h,Q/r_h)$ plane, which is suggestive of several properties of these solutions. 
First, it appears that the map $(\alpha/r_h^2,\beta) \mapsto (M/r_h,Q/r_h)$ is invertible offering a direct connection between the asymptotes of the solution and the underlying gravity model. 
The lines of constant $\beta$ approaching the vertical line in the left figure, merge at $(M/r_h,Q/r_h)=(0.5,0)$, which corresponds to the GR solution.
These same lines appear to be bounded from above as they approach the parabola in $(\alpha/r_h^2,\beta)$.

\begin{figure*}[t]
    \centering
    \includegraphics[width=.47\linewidth]{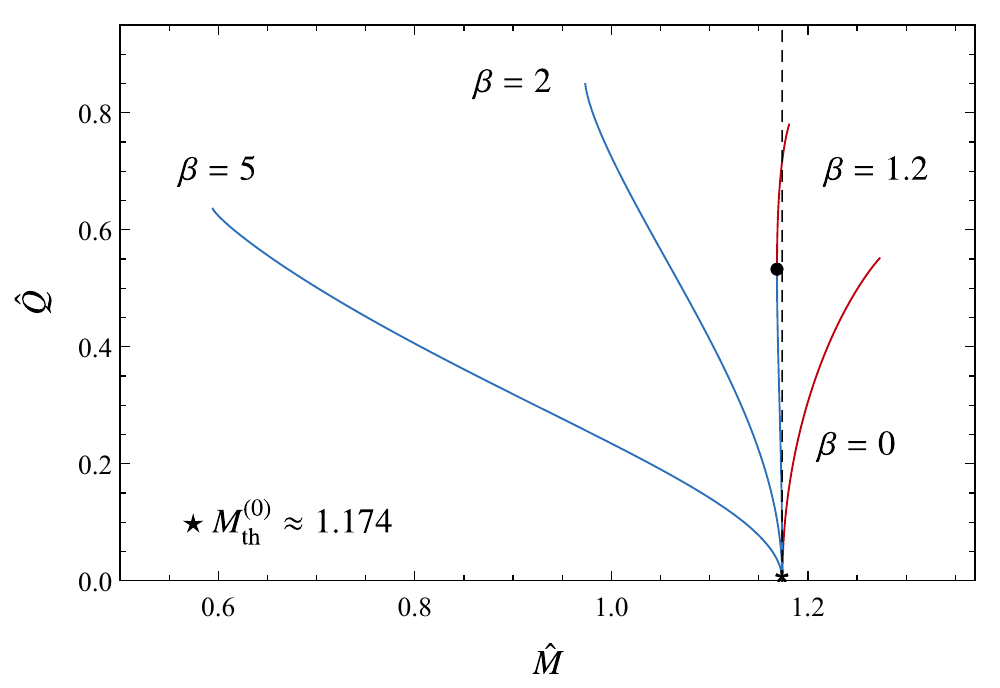}\hfill\includegraphics[width=.47\linewidth]{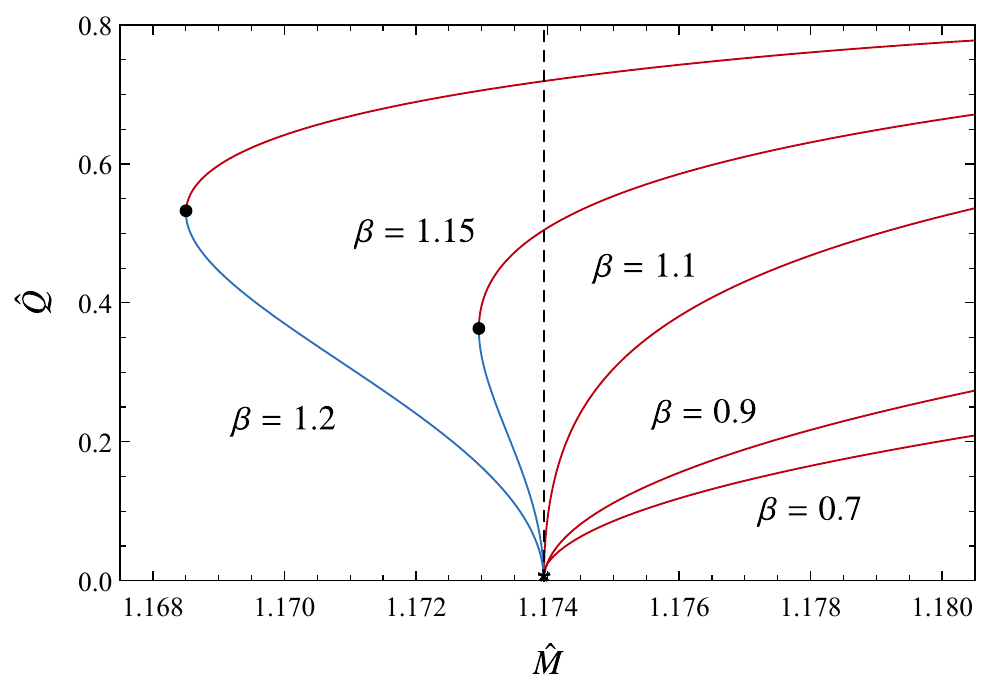}
    \caption{Charge-mass diagram for scalarized BHs with the Ricci term. The blue part of the curves corresponds to stable solutions while the red to unstable ones (as verified by the stability analysis in Sec.~\ref{sec:numerical_results}). The dotted vertical line represents $\hat{M}=\hat{M}_c$. \textit{Left}: Sample of solutions for different values of $\beta$. Solutions on the left part of the dotted vertical line on the diagram are stable against radial perturbations. \textit{Right:} Zoom on the stability threshold region. We see that for $\beta\approx1.15$ stable solutions appear in the theory. For $\beta>\beta_c\approx 1.2$ unstable solutions are no longer present.}
    \label{fig:scalarized_solutions}
\end{figure*}

The BH solutions and their stability are better understood when characterized by the quantities $(\hat M,\hat Q)$~\cite{Macedo:2019sem,Macedo:2020tbm}, as in the rightmost panel of Fig.~\ref{fig:domains}. 
For clarity, in Fig.~\ref{fig:scalarized_solutions} we show the same domain but for select values of $\beta$.

Schwarzschild BHs are radially unstable for $\hat{M}<\hat{M}_c\approx1.174$~\cite{Blazquez-Salcedo:2018jnn,Macedo:2019sem}.
It can be seen that $\hat{M}_c$ is the point where the curves converge in Fig.~\ref{fig:scalarized_solutions}. 
When $\hat{M}<\hat{M}_c$, scalar perturbations grow spontaneously and form a non-trivial scalar profile, and so are the energetically favourable solutions. 
We again stress that the final scalar field profile is determined by the full non-linear equations and not only the terms that trigger the instability.

In contrast, Schwarzschild  BHs are stable in the region $\hat{M}>\hat{M}_c$, and potential scalarized solutions would decay back to GR, as seen from purely energetic arguments. 
This reasoning can be confirmed by performing a radial linear stability analysis of the scalarized BH solutions.

As we increase the value of $\beta$, the curves start to occupy the region $\hat{M}<\hat{M}_c$, and their gradients become negative. 
In accordance with \cite{Macedo:2019sem,Antoniou:2021zoy}, we expect a negative gradient to correspond to radially stable solutions. 
From the right panel of Fig.~\ref{fig:scalarized_solutions}, we see that the critical value of $\beta$ for which part of the curves starts existing in the region $\hat{M}<\hat{M}_c$ is approximately $\beta\approx 1.15$, while the window of potentially stable solutions widens when $\beta$ increases. 
Moreover, beyond $\beta_c\approx 1.2$ we do not find any solutions that we would expect to be radially unstable. 
Note that while the right panel in Fig.~\ref{fig:domains} shows that there are potentially stable solutions with $\beta<0$, it is the $\beta>0$ case that renders GR a cosmic attractor \cite{Antoniou:2020nax}, so we will focus on $\beta>0$ henceforth. 

While the arguments above provide a preliminary indication of the scalarized solutions' stability, it is necessary to investigate the problem by performing a full stability analysis before reaching a conclusion.

\subsection{Radial perturbations}

We can investigate the radial stability of the BH solutions by employing a perturbative approach. 
This not only allows us to further elucidate the timescale of the instability, but also to analyze possible modifications to the oscillatory spectrum of the BHs. 
It is worth noting that since a scalar degree of freedom is absent in GR, the radial modes contribute only to a shift in the mass in that case, and hence are not radiative in nature~\cite{Regge:1957td}.

We start by considering time-dependent radial perturbations of the metric tensor and scalar field over a static and spherically symmetric background:
\begin{align}
    \begin{split}
    ds^2=&-[A_0(r)+A_1(t,r)]dt^2+\frac{dr^2}{B_0(r)+ B_1(t,r)}\\
    &+r^2d\Omega^2,
    \end{split}\\[2mm]
    \phi=&\;\phi_0(r)+\phi_1(t,r)
\end{align}
where $A_0$, $B_0$ and $\phi_0$ are the time-independent background solutions, while $A_1$, $B_2$ and $\phi_1$ are the time-dependent perturbations. We can then write down a system of equations for $(A_1,B_1,\phi_1)$, by substituting the metric and scalar perturbations in Eqs. \eqref{eq:grav_eq} and \eqref{eq:scal_eq}. 
This system can be reduced to a second order partial differential equation system for $\phi_1$~\cite{Blazquez-Salcedo:2018jnn}, namely
\begin{equation}
   g(r)^2\frac{\partial^2\phi_1}{\partial t^2} -\frac{\partial^2\phi_1}{\partial r^2}+C(r)\frac{\partial\phi_1}{\partial r}+U(r)\phi_1=0,
   \label{eq:wavephi}
\end{equation}
where the coefficients depend only on the background solution.

When considering a quadratic coupling function with the Gauss-Bonnet term, Ref.~\cite{Blazquez-Salcedo:2018jnn} pointed out that for some values of the coupling, the equation describing the perturbations is not hyperbolic. While this can hinder the investigation of linear stability as a Cauchy problem, we can still examine the mode structure of the spacetime by looking into the frequencies $\omega$.

To perform a mode analysis of the spacetime, we search for the natural frequencies of the system $\omega$, such that $\phi_1(t,r)=\phi_1(r)e^{-i\omega t}$. The perturbation equation~\eqref{eq:wavephi} can be manipulated to the more familiar Schr\"odinger form:

\begin{equation}
\label{eq:pertEq}
    \left(-\frac{d^2}{d r_*^2}+V_{\text{eff}}\right)\psi\,=\,\omega^2\psi,
\end{equation}
where $\phi_1(r)=F(r)\psi(r)$ and the tortoise coordinate is defined through $ dr_*=g(r)\, dr$. We also define
\begin{align}\label{eq:potential}
    \frac{2F'}{F}&=\;C-\frac{g'}{g},\\
    V_{\text{eff}}&=\;\frac{1}{g^2}\bigg[U+\frac{C^2}{4}-\frac{C'}{2}-\frac{3\,g'^2}{4\, g^2}+\frac{g''}{2g}\bigg].
\end{align}

The real part of the frequency $\omega$ describes the resonant modes of the system, i.e., for which frequencies an initial perturbation would respond. 
The imaginary part of the frequency indicates the system's (modal) stability. 
For modes with a negative imaginary part, an initial perturbation decays exponentially, while when positive the perturbation grows and the system is rendered unstable. 
Generally, the effective potential identified from Eq.~\eqref{eq:potential} is useful when one attempts to verify the presence of unstable modes as a negative value for the integral of the effective potential with respect to the tortoise coordinate indicates the existence of unstable modes \cite{doi:10.1119/1.17935}, namely
\begin{equation}\label{eq:condition}
    \int_{-\infty}^{+\infty}V_{\text{eff}}(r_*)\,dr_*<0 \Rightarrow \; \textit{unstable modes}.
\end{equation}
It is worth noting that while the condition \eqref{eq:condition} is sufficient in indicating the presence of unstable modes, it is not a necessary condition. 

In what follows we will be making use of the compactified coordinate
\begin{equation}
    x=1-r_h/r,
\end{equation}
which maps all of spacetime, from the black hole horizon up to infinity, to a finite region, namely $x \in [0,1]$.

\begin{figure}
    \centering
    \includegraphics[width=\linewidth]{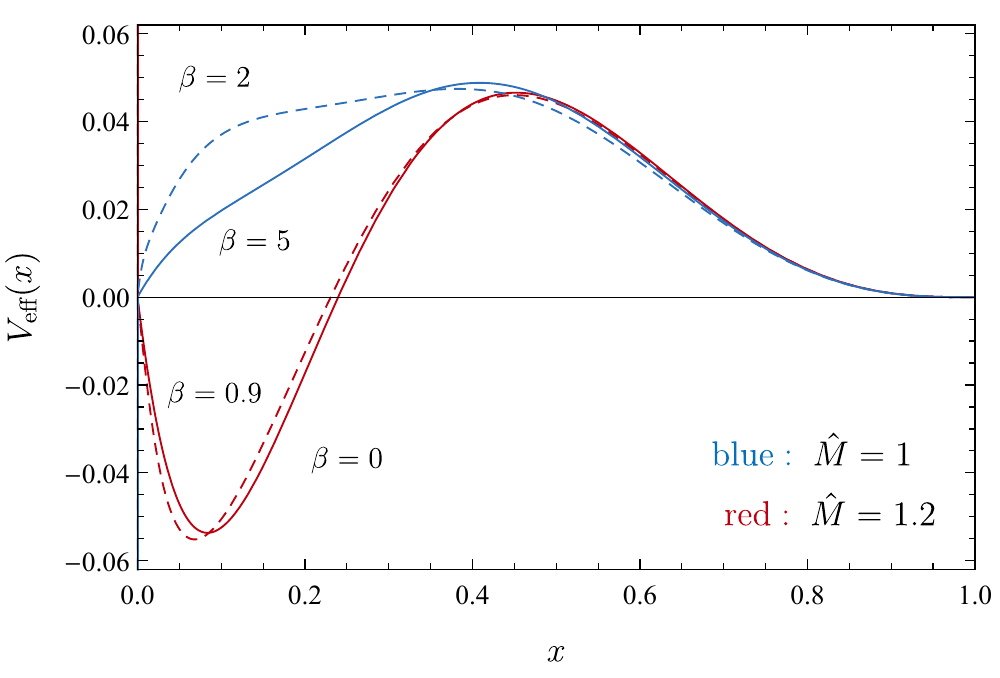}
    \caption{Plot of the effective potential for four different cases. The red lines correspond to right tilting curves with $\hat{M}=1.2$, while the blue ones to left-tilting curves with $\hat{M}=1$. In either case the dashed lines are characterized by a larger value of beta in comparison with the solid ones.}
    \label{fig:potential}
\end{figure}

In Fig.~\ref{fig:potential} we plot the effective potential for two random normalized masses $\hat{M}$ corresponding to the left and right parts of the left panel of Fig.~\ref{fig:scalarized_solutions}, i.e. $\hat{M}=1$ and $\hat{M}=1.2$. 
We see that for $\hat{M}=1.2>\hat{M}_c$ the effective potential has a large negative region which is entirely absent for $\hat{M}=1<\hat{M}_c$. 
Taking into account \eqref{eq:condition}, we have a further indication suggesting the presence of unstable modes for $\hat{M}>\hat{M}_c$ and the absence of them for $\hat{M}<\hat{M}_c$.

To explore the modal structure of the spacetime, we have to impose proper boundary conditions in order to obtain the modes. These correspond to an ingoing wave at the horizon and an outgoing one at infinity
\begin{equation}
\hspace{2.5cm}\Bigg\{\hspace{-2.7cm}
    \begin{split}
        &\phi_1\xrightarrow[r_*\rightarrow -\infty]{x\rightarrow\, 0} e^{-i\omega r_*}\\
        &\phi_1\xrightarrow[r_*\rightarrow +\infty]{x\rightarrow\, 1} e^{+i\omega r_*}
    \end{split}\;.
\end{equation}
We can see see that, for modes with $\omega_I>0$ (unstable), they simplify to $\phi_1(x\rightarrow 0,1)=0$.

\begin{figure}
    \centering
    \includegraphics[width=\linewidth]{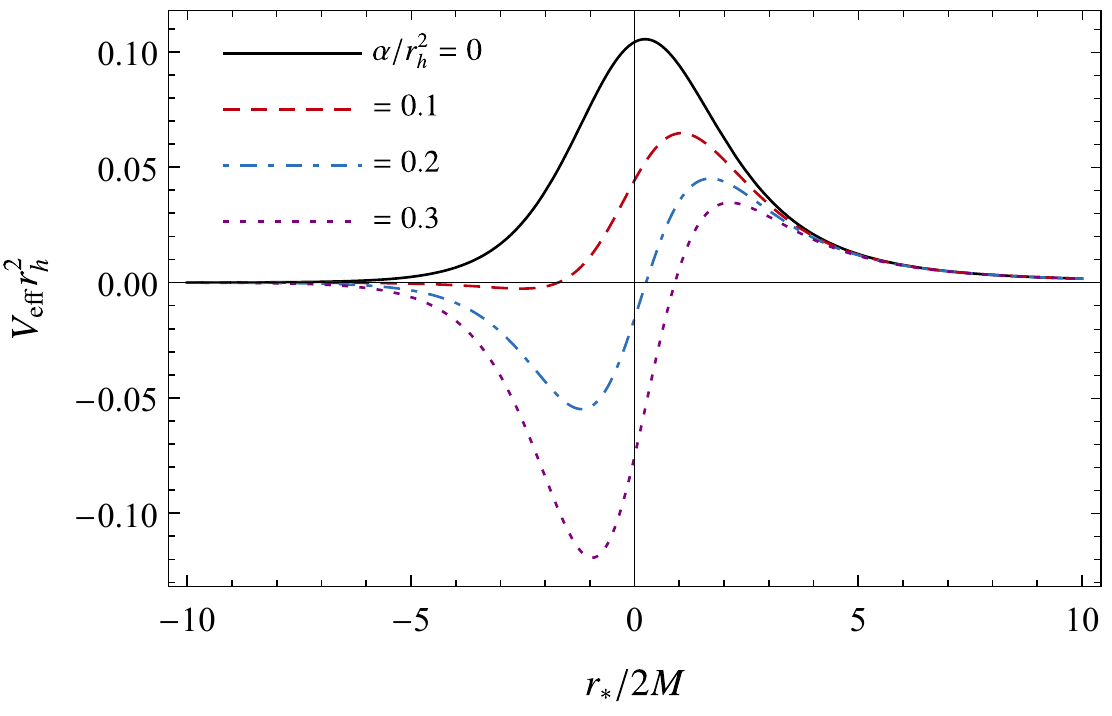}
    \centering\caption{Effective potential for radial perturbations in the Schwarzschild spacetime. The effective potential for the scalarized BH behaves similarly, as can be seen in Fig.~\ref{fig:potential}, presenting a minimum and a maximum as well.}
    \label{fig:potential_sch}
\end{figure}

To finish this section, let us mention that the equations describing the radial perturbations in the Schwarzschild spacetime can be directly obtained from~\eqref{eq:wavephi} by setting $\phi_0=0$ and $A=B=1-r_h/r$ which specify $g$, $C$ and $U$. 
The resulting scalar wave equation is
\begin{equation}
    \frac{\partial^2\phi_1}{\partial t^2}-\frac{\partial^2\phi_1}{\partial r_*^2}+\left(1-\frac{r_h}{r}\right) \left(\frac{r_h}{r^3}-12 \frac{\alpha\, r_h^2}{r^6}\right)\phi_1=0,\label{eq:waveschwarzs}
\end{equation}
where $dr_*=dr/(1-r/r_h)$ is the tortoise coordinate of the Schwarzschild spacetime. The effective potential is identified as
\begin{equation}
    V_{\text{eff}}^{(\text{d})}=\left(1-\frac{r_h}{r}\right)\left(\frac{r_h}{r^3} -\frac{12\, r_h^2\alpha}{r^6}\right),
\end{equation}
where the index \textit{d} stands for \textit{decoupling}. 
In Fig.~\ref{fig:potential_sch} we plot the potential in the decoupling limit for some values of $\alpha$, using the tortoise coordinate to improve visualization. 
Just as before, the equation describing radial perturbations can be used to access the stability properties of the Schwarzschild spacetime in sGB. 
From a direct integration of the potential, it is straightforward to see that \eqref{eq:condition} in this case yields $\alpha/r_h^2\gtrsim 0.208$.

\section{Numerical results}\label{sec:numerical_results}
In order to find the modes of scalarized BHs, we follow the direct integration method presented in previous works on the same subject~\cite{Blazquez-Salcedo:2018jnn,Silva:2018qhn}. 
We briefly summarize the method here.

After picking a value for $\omega$, we integrate Eq.~\eqref{eq:wavephi} from the horizon and infinity, using in-going and-outgoing waves as boundary conditions respectively\footnote{We could instead work with Eq.~\eqref{eq:pertEq}, but in our setup this would add an extra step, slowing down the integrations.}. 
In practice, the integration starts from finite values very close and very far away from the horizon's position, such that the potential is small. 
This changes the boundary conditions which are no longer purely in-going and out-going waves, but are rather given by the Taylor expansion of the field at the horizon and infinity.
Using the two separate solutions, we can demand that they are linearly dependent on a given frequency $\omega$. This is done by examining the Wronskian, given by
\begin{equation}
    W =\bigg[ \phi_1^{(-)}\frac{\partial \phi_1^{(+)}}{\partial r_*} - \phi_1^{(+)}\frac{\partial \phi_1^{(-)}}{\partial r_*}\bigg]\,.
\end{equation}
where $\phi_1^{(-)}$ represents the solution obtained by integrating from the horizon and $\phi_1^{(+)}$ the solution obtained from infinity. The Wronskian vanishes when the value of $\omega$ is a QNM frequency.
An alternative approach is to integrate from the horizon using the in-going boundary condition, up to a large distance $r_{\infty}/r_h$. 
Then one can decompose the solution at infinity onto in-going and out-going waves, and should the value of $\omega$ be a QNM frequency, the in-going amplitude is zero. 
For both methods, an initial choice for $\omega$ is made, and root-finding algorithms are employed to solve for the QNM frequency. 
This is usually called the shooting method, as one starts at one end, ``shooting'' for the value of $\omega$ for which the boundary condition is satisfied at the other end.

Both methods are suitable for finding stable quasinormal modes with large quality factors, i.e., with large $|\omega_r/\omega_i|$. This usually means that the fundamental mode is easily found. 
The reason is that the function describing radial perturbations grows exponentially in $r$, as $\sim e^{\omega_i r_*}$. 
The method also works remarkably well with unstable modes, as the perturbations decrease exponentially with $r$. 
As such, this method is reliable in identifying regions where BHs are linearly unstable.

BHs solutions of the theory~\eqref{eq:ActionGeneric} were already studied to some extent in Ref.~\cite{Antoniou:2021zoy} and we revisited some in Figs.~\ref{fig:domains} and~\ref{fig:scalarized_solutions}. 
Now we present new insights considering the stability and modes of the Schwarzschild and scalarized BHs.

\subsection{Radial oscillations and the existence of purely imaginary modes}

While the Schwarzschild BH is a solution in both GR and sGB theories, its dynamical response to perturbations can be completely different. In fact, it is precisely this difference that allows for spontaneous scalarized BHs, where the radial perturbations instabilities lead to the scalar hair. 
Here we investigate the radial mode structure of BHs in sGB, elucidating some major points considering Schwarzschild and scalarized solutions.

\subsubsection{Scalarized BHs}
\begin{figure}
    \centering
    \includegraphics[width=\linewidth]{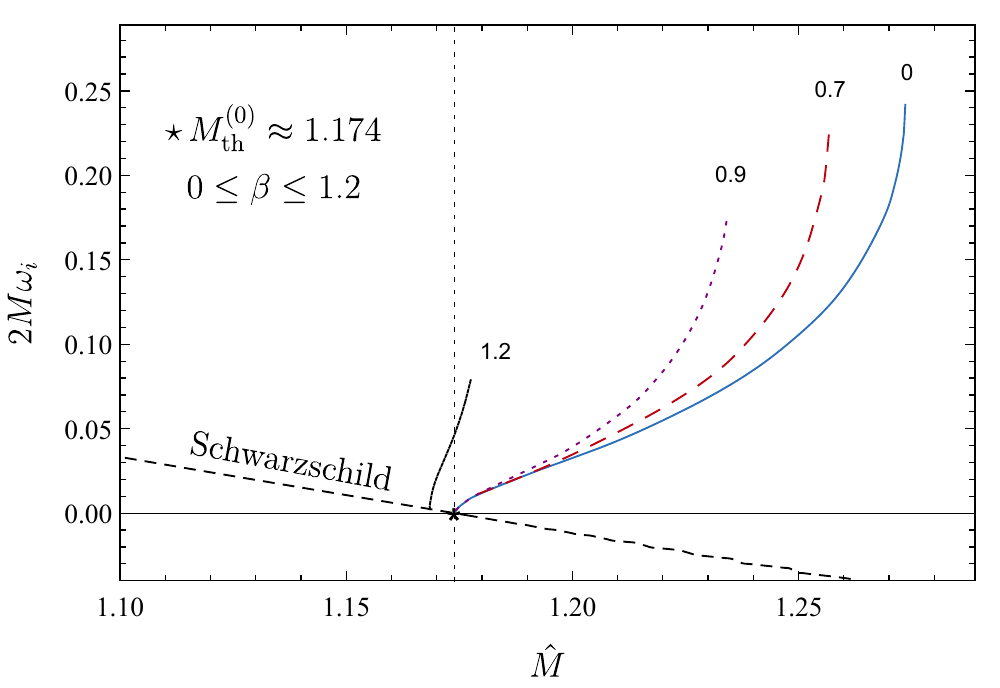}
    \caption{Imaginary part of the unstable modes for scalarized solutions in sGB considering the Ricci term. We see that as we increase the value of $\beta$, the instability decreases, having a critical value $\beta_c< 1.2$ for which the $\hat{M}$-$\hat{Q}$ curves start tilting to the left and solutions becomes stable. We found no unstable modes for scalarized BHs with $\beta>\beta_c$. The vertical dotted line marks the scalarization threshold and the dashed black line the Schwarzschild  mode.}
    \label{fig:modes_scalarized}
\end{figure}

Let us look into the solutions presented in Fig.~\ref{fig:scalarized_solutions}. 
As discussed, we expect to find unstable modes for the region $\hat{M}>\hat{M}_c$, where the Schwarzschild BH is stable and energetically favourable. 
We performed a search for the modes using the shooting method. 
In Fig.~\ref{fig:modes_scalarized} we plot the unstable mode frequencies for the scalarized solutions considering different values for $\beta$, some of them presented in the right panel of Fig.~\ref{fig:scalarized_solutions}. These frequencies are purely imaginary. 
The imaginary part of the Schwarzschild fundamental mode is also plotted, and it is independent of our choice of $\beta$. 
We notice that all scalarized solutions with $\hat{M}>\hat{M}_c\approx1.175$ are unstable.

In agreement with the predictions made by observing Fig.~\ref{fig:scalarized_solutions}, for $\beta>\beta_c\approx 1.15$, the behavior of the curves changes. 
We notice that the curve for which $\beta=1.2$ begins not from $\hat{M}_c$, but from some $\hat{M}\approx1.168<\hat{M}_c$. 
This means that the minimum BH mass for this parameter is no longer $\hat{M}_c$, but smaller.  
Once again, the reasoning for this can be understood qualitatively from purely energetic arguments. 
For $\beta=1.2$, we can have two scalarized solutions for a given mass $\hat{M}$, presenting different charges. 
Solutions with higher charges are unstable, decaying to the scalarized BH with a smaller charge. 
A similar feature was observed for the case of scalarized BHs with self-interaction~\cite{Macedo:2019sem}. 
Overall, the timescale of the instability $\tau=|\omega_I^{-1}|$ increases as $\beta$ increases, indicating the shift from unstable to stable solutions.

\subsubsection{Radial modes for the Schwarzschild spacetime}

Since Schwarzschild BHs are stable for $\hat M>\hat M_c$ and the dynamical response is different from GR, it is natural to investigate the impact of sGB terms on the Schwarzschild BH spectrum. The fundamental mode was already analyzed in Ref.~\cite{Macedo:2020tbm}, elucidating how the transition from stable to unstable modes occurs. Here we take an additional step, looking into the first overtone. The results of this subsection are independent of $\beta$ [cf. Eq.~\eqref{eq:waveschwarzs}].

\begin{figure*}
    \centering
    \includegraphics[width=.47\linewidth]{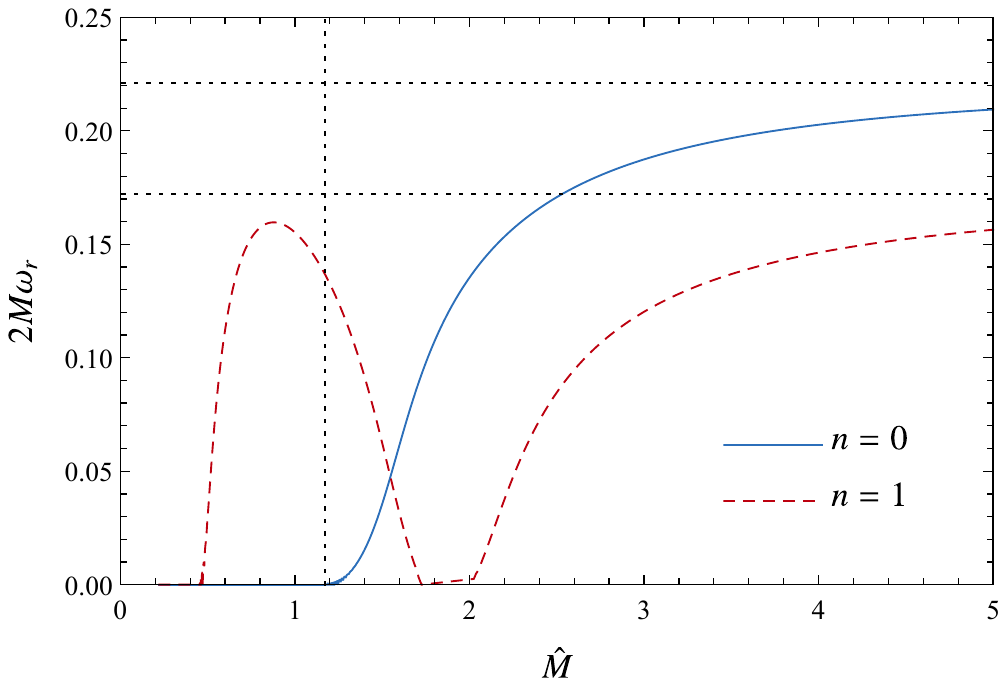}\hfill\includegraphics[width=.47\linewidth]{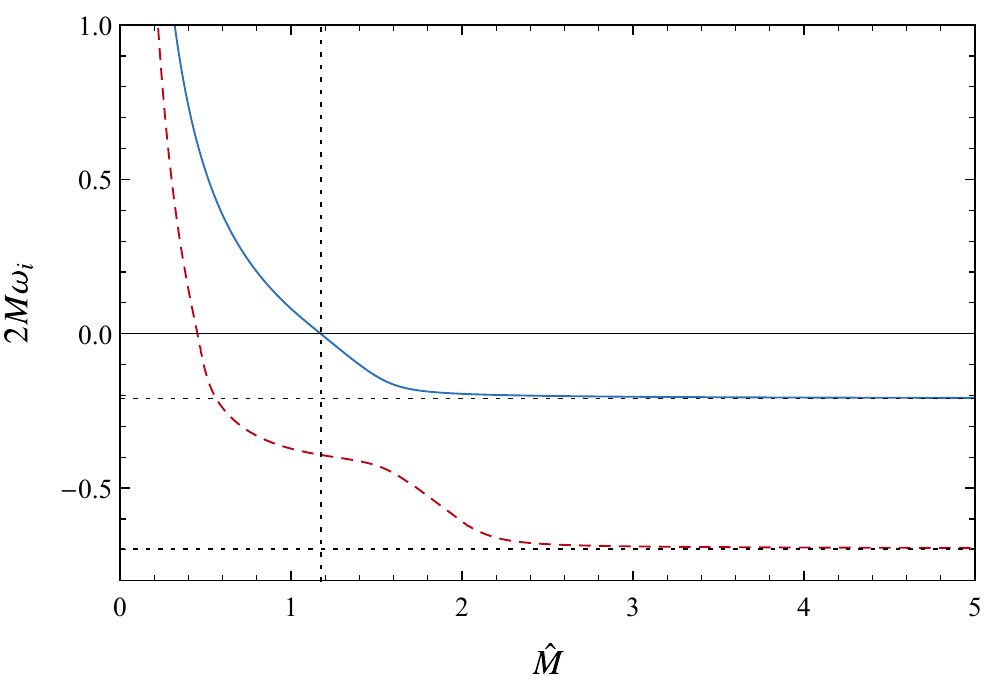}
    \caption{Fundamental mode and first overtone of radial QNM frequencies in the Schwarzschild spacetime. We see that the fundamental mode is responsible for the instability. The first overtone presents a purely imaginary mode with $\omega r_h\approx -i0.55$ for $\hat{M}\approx 1.87$ in a region that the BH is \textit{stable}. Higher overtones present the same feature. The horizontal dotted lines represent the GR scalar modes and the vertical line, the threshold for scalarization.}
    \label{fig:mode_sch}
\end{figure*}

To find the modes for the Schwarzschild BH we use the \textit{continued fraction} (CF) method, as presented in Appendix~\ref{app:cf} (see also Ref.~\cite{Macedo:2020tbm}). In Fig.~\ref{fig:mode_sch} we show the results. Starting from the rightmost part of the plots, we see that for $\hat{M}\gg 1$ (or equivalently $\alpha^2\ll M$) we recover the modes for a minimally coupled scalar field in GR, as expected. As we decrease $\hat{M}$, both the real and the imaginary part of the fundamental mode ($n=0$) approach zero monotonically at $\hat{M}=\hat{M}_c$. Beyond that point the frequency becomes purely imaginary and the corresponding mode unstable. For the first overtone ($n=1$), however, we notice that the real part seems to approach zero \textit{before} the instability occurs. This would imply that Schwarzschild BHs with $\hat{M}\approx 1.87$ in this theory have a purely imaginary mode, with the first one being $\omega r_h\approx -i0.55$. The real part of the first overtone goes to zero again when this mode becomes unstable (purely imaginary), in the region $\hat M<\hat M_c$. Higher overtones present a similar behavior, suggesting that we might have many of these purely imaginary modes even for stable Schwarzschild BHs in sGB theories. 

We highlight here that the usual CF method is not suitable to study purely imaginary modes~\cite{Cook:2016fge,Cook:2016ngj}. As the real part of the modes decreases, more terms in the CF expansion are needed in order to compute reliable values for the QNM frequencies. Typically, near the purely imaginary mode, we consider $N=5\times 10^4$ or more terms in the CF expansion. Curiously, only recently this class of modes was computed with a reliable precision for the Kerr spacetime~\cite{Cook:2016fge,Cook:2016ngj}. We shall not attempt to generalize this method for sGB theories, but it would be interesting to further investigate the full spectrum of the Schwarzschild spacetime in the theory in light of such tools.

Purely imaginary modes for BHs are not an exclusive feature of sGB theories. As mentioned above, even within GR, the existence of such a class of modes has been known for quite some time. For instance, Schwarzschild BHs in GR have algebraically special modes that are purely imaginary~\cite{Chandrasekhar:1975nkd}, and many works have studied these modes for Kerr BHs. While the physical implications of these purely imaginary modes in BH spacetimes are still poorly understood, it is interesting to see them arising in the first overtone for Schwarzschild BHs in sGB. This feature, which seems to be an exception for GR BHs, seems to be common in BHs in sGB.

\subsection{The hyperbolic nature of the equations}

An important property of partial differential equations pertains to hyperbolicity. In physical theories, hyperbolic equations are necessary to describe the time evolution of initial data. In GR, where equations are quasilinear, proof of strong hyperbolicity (e.g. through the \textit{harmonic gauge} choice) establishes the well-posedness of the theory (see Ref.~\cite{Wald:1984rg} for a pedagogical discussion). In linear equations such as \eqref{eq:wavephi}, the situation is fairly straightforward. Eq. \eqref{eq:wavephi} is hyperbolic provided that $g^2(x)>0$. It has been shown that for the exponential coupling, the equation describing the radial perturbations is not hyperbolic for a variety of solutions~\cite{Blazquez-Salcedo:2018jnn}. Interestingly, even though the equations are not hyperbolic, we can still proceed to search for unstable modes. It was shown that for scalarized BHs in the exponential model with low mass, an unstable mode arises in the region where hyperbolicity is broken. Here we investigate the impact of the Ricci term on the hyperbolicity of the equation describing radial perturbations.

We start by analyzing the behavior of the coefficient $g(r)^2$ in the near-horizon regime. In this section, we shall replace the quadratic coupling with the Gauss-Bonnet term by a generic function of the form $\alpha \phi^2\to\alpha f(\phi)$, in order to compare our results with other works in the literature.
Note that the definition of the normalised charge and mass are unchained under this substitution.
Using the expansion of the background near the event horizon we find that  the coefficient $g(r)^2$ appearing in Eq.~\eqref{eq:wavephi} behaves as
\begin{equation}
    g(r)^2(r-r_h)^2\approx  \frac{1}{2 a_1}\left(1+\frac{\sqrt{\delta}}{\gamma}\right),
\end{equation}
where
\begin{align}
    \delta&=73728 \alpha ^3 \beta  \phi_h f'\left(\phi_h\right)^3\nonumber\\
    &-768 \alpha ^2 \left[\beta  (9 \beta -2) \phi_h^2+8\right] f'\left(\phi_h\right)^2\nonumber\\
    &+\left[\beta  (3 \beta -1) \phi_h^2+4\right]^2,\\
    \gamma&=\beta  \phi_h \left[(3 \beta -1) \phi_h-48 \alpha  f'\left(\phi_h\right)\right]+4.
\end{align}
In order to investigate whether the Ricci term helps maintaining the hyperbolic nature of the equation, we look into large positive values of $\beta$ that, from our previous analysis, we know correspond to stable scalarized solutions. We find
\begin{align}
    g(r)^2(r-r_h)^2\sim&\frac{1}{a_1}\bigg[1+\frac{8\alpha}{\phi_h\beta}+\frac{8\alpha f'(\phi_h)}{3\phi_h^2\beta^2}(\phi_h-24\alpha)\nonumber\\
    &+{\cal O}(\beta^{-3})\bigg],
    \label{eq:hyperbolic_c}
\end{align}
which indicates that the Ricci term in the action acts \textit{in favor} of the hyperbolic character of the equation. Note, however, that in order to verify whether the equation maintains its hyperbolicity we need to properly solve the BH solution and obtain the coefficient $g(r)^2$ numerically.

\begin{figure}
    \centering
    \includegraphics[width=1\linewidth]{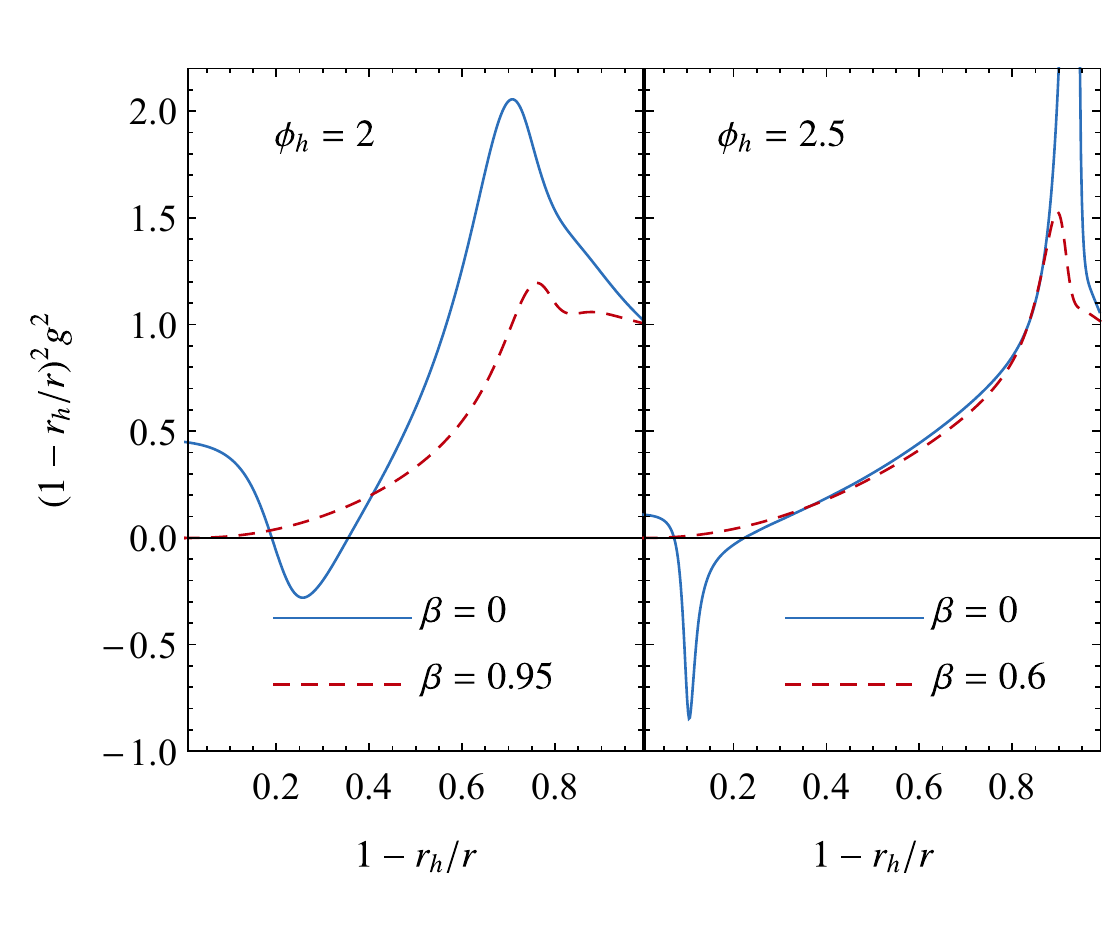}
    \caption{Respresentative solutions illustrating the effect of hyperbolicity healing from the Ricci term. The equations are not hyperbolic in the region $g^2<0$. The solid lines have $(\hat{M},\hat{Q})=(0.2677,0.4494)$ (left panel) and $(\hat{M},\hat{Q})=(0.0319, 0.110)$ (right panel). The dashed lines have $(\hat{M},\hat{Q})=(0.0915,0.195)$ (left panel) and $(\hat{M},\hat{Q})=(0.029, 0.068)$ (right panel).}
    \label{fig:compair_hyper}
\end{figure}
Now, let us consider a scenario that is known to break hyperbolicity and check the influence of the Ricci term. We consider the exponential model presented in Ref.~ \cite{Blazquez-Salcedo:2018jnn}, where 
\begin{equation}
f(\varphi)\propto [1-e^{-3\phi^2/2}],
\end{equation}
looking into solutions with a fixed $\phi_h$ and varying $\beta$. 
In Fig.~\ref{fig:compair_hyper} we compare two scalarized solutions with $\phi_h=2$ and $2.5$. 
We observe that the radial domain in which the perturbation equation is non-hyperbolic decreases as $\beta$ increases, as predicted by Eq.~\eqref{eq:hyperbolic_c}.
Further, we observe that for some limiting value of $\beta$ the region with $g^2<0$ seems to vanish and the solution is hyperbolic for all $r>r_h$. 
We note, however, that as we approach this threshold the solutions for a given $\phi_h$ are increasingly hard to find, and beyond the threshold solutions cease to exist. 
It seems that while the $\beta$-term improves the hyperbolicity of the perturbation equation, it still is not enough to ensure it for all values of $(r,\phi_h)$ for a given $\beta$.

\begin{figure}
    \centering
    \includegraphics[width=\linewidth]{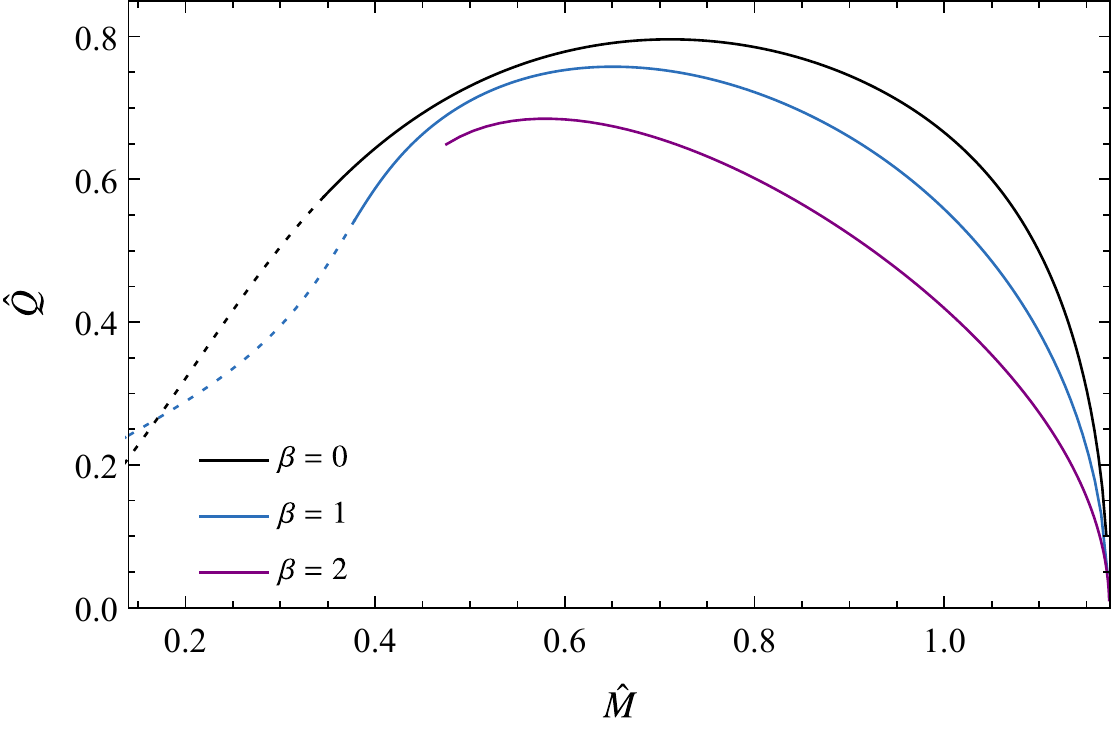}
    \caption{
    Normalised charge-mass diagram for scalarized black holes considering the exponential GB coupling and the Ricci term. 
    The dotted lines indicate the region in which the pertubation equations are non-hyperbolic. 
    For $\beta=2$ the perturbation equation is always hyperbolic.}
    \label{fig:exponential}
\end{figure}

To further illustrate the hyperbolic properties of the equations as a function of the background solution, Fig.~\ref{fig:exponential} shows curves of constant $\beta$ and varying $\phi_h$ in the normalised charge-mass plane for the exponential model.
The dotted part of the curves corresponds to regions in which the radial perturbation equations are not hyperbolic. 
That the $\beta = 2$ curve ends is indicative of our inability to find any solutions past this point. It seems that while the additional term helps with hyperbolicity, the parameter space of the solutions is truncated.
 
\section{Conclusions}\label{sec:conclusions}

We have explored the influence of a coupling between a scalar field and the Ricci scalar on linear perturbations around scalarized black holes. 
This coupling, which is expected to be present in scalarization models based on EFT considerations, has already been shown to be crucial for observational viability: it can make GR a cosmological attractor, thereby providing the right conditions for compact objects without cosmological fine-tuning \cite{Antoniou:2020nax}, and it can also suppress neutron star scalarization, thus removing binary pulsar constraints \cite{Ventagli:2021ubn}. 
It also affects the amount of scalar charge scalarized black holes can carry \cite{Antoniou:2021zoy}. 
We performed a radial stability analysis and have numerically shown that this same term renders scalarized BH solutions stable, confirming the expectations of Ref.~\cite{Antoniou:2021zoy}. 

This happens for values of the coupling constants that are within the same range considered in the aforementioned papers. 
Indeed, we have not found unstable modes for $\beta$ larger than some critical value, $\beta\approx 1.2$. 
Choosing a $\beta$ above this threshold is consistent with having a cosmological attractor \cite{Antoniou:2020nax} and could quench neutron star scalarization for a range of $\alpha$ that still leads to black hole scalarization \cite{Ventagli:2021ubn,Antoniou:2021zoy}. It is worth noting that negative values of $\beta$ are also capable of stabilizing the solutions in a similar manner, but since they do not give rise to the cosmological attractor feature, they seem to be less interesting.

We have also performed a radial mode analysis in the Schwarzschild spacetime and looked for QNM modes. We were able to illustrate an interesting property: Beyond the fundamental mode one can find purely imaginary modes in a region of the parameter space where Schwarzschild BHs are stable.

Finally, we analyzed the effects of the scalar-Ricci coupling on the hyperbolicity of the scalar perturbation equation, using the exponential GB coupling as an example. 
We demonstrated that it actually improves the hyperbolic nature of the problem, by reducing the region of the parameter space where hyperbolicity breaks down. 
This happens as the additional term changes the scalar field profile, as determined by the full non-linear field equations, which in turn changes the coefficients of the linear perturbation equations. It would be interesting to generalize the hyperbolicity analysis to more general perturbations and beyond the linear level, in order to check if the coupling with the Ricci scalar could have a positive effect when considering hair formation by collapse \cite{Ripley:2020vpk} or binary mergers \cite{East:2021bqk}.

\section*{Acknowledgements}
G.A. acknowledges partial financial support from the Onassis Foundation. C.F.B.M. would like to
thank Fundação Amazônia de Amparo a Estudos e Pesquisas (FAPESPA), Conselho
Nacional de Desenvolvimento Científico e Tecnológico (CNPq), and Coordenação de
Aperfeiçoamento de Pessoal de Nível Superior (CAPES), from Brazil. 
R.M. acknowledges support from the STFC Consolidated grant ST/T000732/1.
T.P.S. acknowledges partial support from the STFC Consolidated Grants no. ST/T000732/1 and no. ST/V005596/1.

\appendix

\section{Background field equations}\label{app:equations}
From the nontrivial components of the zeroth-order Einstein
equations, we obtain

\begin{widetext}
    \begin{align}
    \begin{split}
        (t,t):\quad&B \left\{\beta  \phi ^2+2 \phi  \left[\phi '' \left(-8 \alpha +8 \alpha  B+\beta  r^2\right)+2 \beta  r \phi '\right]+\phi'^2\left[-16 \alpha +16 \alpha  B+(2 \beta -1) r^2\right]-4\right\}+4=0\\
        &-\beta \, \phi ^2+B' \left[\phi \, \phi ' \left(-8 \alpha +24 \alpha B+\beta  r^2\right)+\beta  r \phi ^2-4 r\right]=0,
    \end{split}\\[2mm]
    \begin{split}
        (r,r):\quad& B \phi  \, \phi ' \left[A' \left(-8 \alpha +24 \alpha  B+\beta  r^2\right)+4 A \beta  r\right]+\left(\beta  \phi ^2-4\right) \left[B r A'+A (B-1)\right]+A B r^2 \phi'^2=0,
    \end{split}\\[2mm]
    \begin{split}
        (\theta,\theta):\quad&A \,A' \left\{B' \left[48 \alpha  B \phi  \phi'+r \left(\beta  \phi ^2-4\right)\right]+2 B \left(\beta  \phi ^2+16 \alpha  B \phi'^2+2 \beta  r \phi  \phi'-4\right)\right\}\\
        &+2 A^2 B' \left[\beta  \phi  \left(2 r \phi '+\phi \right)-4\right]+8 A B \phi\,  \phi'' \left(4 \alpha  B A'+A \beta  r\right)+2 A B A'' \left[16 \alpha B \phi \, \phi '+r \left(\beta  \phi ^2-4\right)\right]\\
        &-B A'^2 \left[16 \alpha  B \phi \,\phi'+r \left(\beta  \phi ^2-4\right)\right]+4 A^2 B \phi' \left[2 \beta  \phi +(2\beta -1) r \phi'\right]=0,
    \end{split}\\[2mm]
\end{align}
where a prime indicates differentiation with respect to r. The background equation for the scalar field reads

\begin{equation}
    \begin{split}
        &\phi  \left\{A A' \left[B' \left(24\alpha B -8\alpha+\beta  r^2\right)+4 \beta  B r\right]+4 A^2 \beta  \left(r B'+B-1\right)-B A'^2 \left(8 \alpha  (B-1)+\beta  r^2\right)\right\}\\
        &+2 A r \phi ' \left(B r A'+A r B'+4 A B\right)+2 A B \phi  A'' \left[8\alpha (B-1)+\beta  r^2\right]+4 A^2 B r^2 \phi ''=0.
    \end{split}
\end{equation}
\end{widetext}

\section{Continued fraction method for Schwarzschild BHs in sGB theories}\label{app:cf}

The direct integration method has low accuracy for modes with low quality factors. For these modes, which can be expected at the onset of scalarization, we can implement the \textit{continued fraction} (CF) method to understand how the Schwarzschild spacetime becomes unstable. The CF method relies on providing a semi-analytical approximation of the wave function through the Frobenius method~\cite{Leaver:1990zz}. In summary, the solution can be written in terms of coefficients that are computed by a recursive relation that takes the form of a continued fraction, justifying the name. Since the method requires the analytical form of the coefficients (at least for the expansion to be implemented), we focus mostly on the stability analysis of Schwarzschild BHs.

We look into Eq.~\eqref{eq:waveschwarzs} factorizing the wave-function in the following way~\cite{Macedo:2020tbm}
\begin{align}
    \phi_1(t,r)=&e^{-i\omega t}\left(\frac{r}{r_h}-1\right)^{-ir_h\omega}\left(\frac{r}{r_h}\right)^{2ir_h\omega}e^{i\omega r}\times\nonumber\\
    &\sum_{n=0}^{\infty}a_n(1-r_h/r)^n.
\end{align}
By substituting the above expansion into Eq.~\eqref{eq:waveschwarzs}, combining and simplifying the coefficients, we obtain the following six-term recurrence relation for $a_n$
\begin{widetext}
\begin{equation}
\begin{array}{lr}
\alpha_0 a_{1}+\beta_0 a_0=0,&n=0\\
\alpha_1 a_{2}+\beta_1 a_1 +\gamma_1 a_{0}=0,&n=1\\
\alpha_2 a_{3}+\beta_2 a_2 +\gamma_2 a_{1}+\delta_2 a_{0}=0,&n=2\\
\alpha_3 a_{4}+\beta_3 a_3 +\gamma_3 a_{2}+\delta_3  a_{1}+\sigma_3 a_{0}=0,&n=3\\
\alpha_n a_{n+1}+\beta_n a_n +\gamma_n a_{n-1}+\delta_n a_{n-2}+\sigma_n a_{n-3}+\theta_n a_{n-4}=0, &n>3,
\end{array}
\end{equation}
where
\begin{align}
\alpha_n&=(n+1) r_h^2 (n-2 i r_h \omega +1),\\
\beta_n&=12 \alpha -r_h^2 \left[2 n (n+1)+1\right]+4 i r_h^3 (2 n \omega +\omega )+8 r_h^4 \omega ^2,\\
\gamma_n&=-4 \left[12 \alpha -\frac{1}{4} r_h^2 (n-2 i r_h \omega )^2\right],~
\delta_n=72 \alpha,~~ \sigma_n=-48 \alpha, ~~ \theta_n=12 \alpha.
\end{align}
\end{widetext}
There are different methods used to solve the above equations. Usually, in BH mode analysis, one reduces the $n$-term recurrence relation to a three-term one by successive Gaussian elimination steps, finding the appropriate CF expression to obtain the mode~\cite{Leaver:1990zz,Pani:2013pma}. For example, in the six-term recurrence relation, we have to use three Gaussian elimination steps. We shall use the above to compute the modes for the Schwarzschild spacetime in sGB theories. We usually truncate the series from a higher value for $n=N$ moving backwards up to $n=1$, solving the series~\cite{Pani:2013pma}. The number $N$ of necessary terms depends on the quality factor of the modes, beginning higher for low values of lower quality factors. We note that for $\alpha=0$ we recover the standard recurrence relation for a free scalar field in the Schwarzschild  spacetime. 

\newpage
\bibliography{bibnote}

\end{document}